\documentclass[%
 reprint,
 amsmath,amssymb,
 aps,
pra,
]{revtex4-1}

\usepackage{graphicx}
\usepackage{dcolumn}
\usepackage{bm}
\usepackage{braket}
\usepackage{amsmath}

\begin{document}

\preprint{APS/123-QED}

\title{Ultra-broadband Entangled Photons on a Nanophotonic Chip}

\author{Usman A. Javid$^{1,\uparrow}$}
\author{Jingwei Ling$^{2,\uparrow}$}%
\author{Jeremy Staffa$^{1}$}%
\author{Mingxiao Li$^{2}$}%
\author{Yang He$^{2}$}%
\author{Qiang Lin$^{1,2}$}%
\email{qiang.lin@rochester.edu}
\affiliation{$^{1}$Institute Of Optics, University of Rochester, Rochester NY 14627, USA}
\affiliation{$^{2}$Department of Electrical and Computer Engineering, University of Rochester, Rochester NY 14627, USA}
\affiliation{$^{\uparrow}$These authors contributed equally.}
\date{\today}

\begin{abstract}
Nanophotonic entangled-photon sources are a critical building block of chip-scale quantum photonic architecture and have seen significant development over the past two decades. These sources generate photon pairs that typically span over a narrow frequency bandwidth. Generating entanglement over a wide spectral region has proven to be useful in a wide variety of applications including quantum metrology, spectroscopy and sensing, and optical communication. However, generation of broadband photon pairs with temporal coherence approaching an optical cycle on a chip is yet to be seen. Here we demonstrate generation of ultra-broadband entangled photons using spontaneous parametric down-conversion in a periodically-poled lithium niobate nanophotonic waveguide. We employ dispersion engineering to achieve a bandwidth of 100 THz (1.2 \--- 2 $\mu$m), at a high efficiency of 13 GHz/mW. The photons show strong temporal correlations and purity with the coincidence-to-accidental ratio exceeding $10^5$ and $>$ 98\% two-photon interference visibility. These properties together with the piezo-electric and electro-optic control and reconfigurability, make thin-film lithium niobate an excellent platform for a controllable entanglement source for quantum communication and computing, and open a path towards femtosecond metrology and spectroscopy with non-classical light on a nanophotonic chip.
\end{abstract}

\maketitle


\section{\label{sec:level1}Introduction}

\begin{figure*}
\centering
  \includegraphics[scale=0.45]{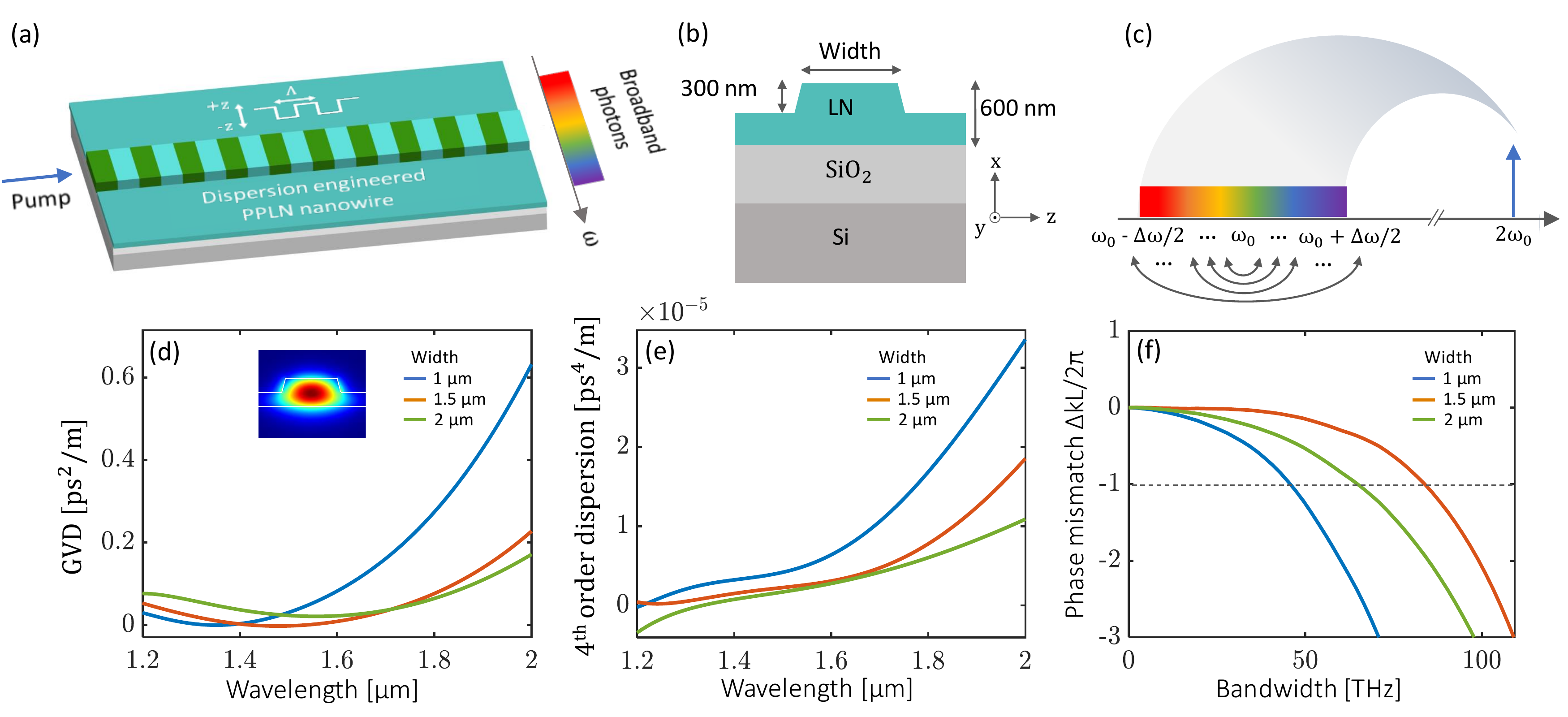}
  \caption{Broadband spontaneous parametric down-conversion (SPDC). (a) A periodically poled lithium niobate (PPLN) on insulator waveguide is dispersion engineered to generate broadband photons with the device cross-section shown in (b). The dispersion engineered device generates a broadband SPDC spectrum with pump at frequency $2\omega_0$ creating two photons with frequencies equally spaced around the center frequency $\omega_0$. Controlling dispersion with waveguide width in the telecom band with variations in the zero dispersion wavelength shown in (d) and the fourth-order dispersion plotted in (e). The inset in (d) shows a cross-section of the waveguide with the fundamental quasi-TE mode at 1550 nm. (f) the phase-mismatch $\mathrm{\Delta k}L$ corresponding to each width is plotted for a waveguide length L = 5 mm with an appropriate choice of poling period $\Lambda$. Dashed line indicates the point where the phase-mismatch Sinc function reaches zero.}
  \label{fig1}
\end{figure*}

\begin{figure*}
\centering
  \includegraphics[scale=0.9]{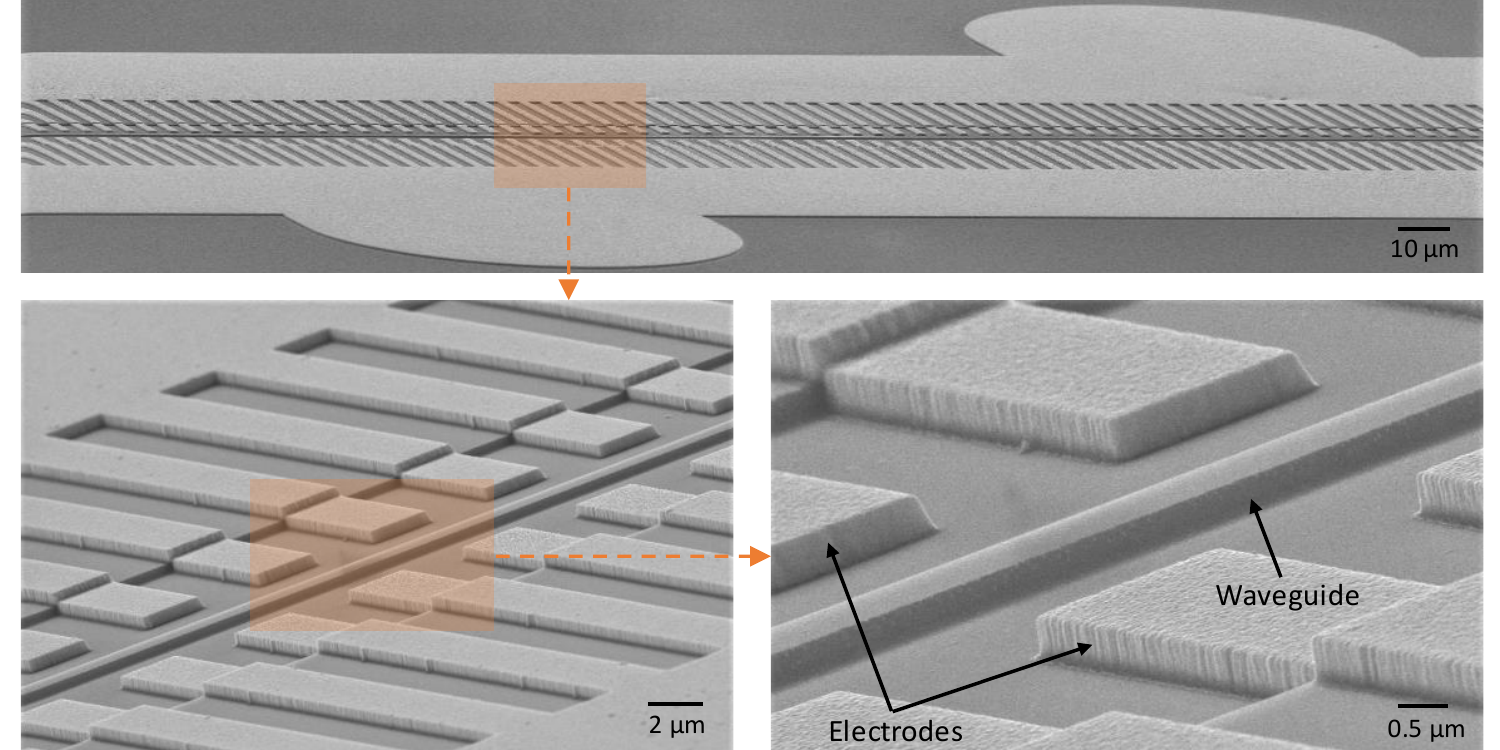}
  \caption{Scanning electron microscope (SEM) images of a fabricated device at different magnifications.}
  \label{figSEM}
\end{figure*}

Photonic quantum technologies have seen significant advances in the past decade with great promise for broad applications in secure communication, metrology and sensing, and advanced computing \cite{photon_int}. Photonic entanglement is a crucial ingredient in these applications. In recent years, significant efforts have been devoted to developing entangled photons on a variety of chip-scale platforms \cite{Review_photon1,Review_photon2}, where the strong light-matter interaction and the engineering flexibility of photonic integrated circuits show great potential for significantly improving the performance and resource requirements for complex functionalities. 

A broad spectral width of quantum entanglement, with an ultra-short coherent time 
would add significant advantages to quantum photonic applications, such as enhancing sensitivity and/or resolution in metrology \cite{Qmetr}, lithography \cite{litho}, spectroscopy \cite{Qspec}, 
nonlinear microscopy \cite{TPAmicroscope}, quantum optical coherence tomography \cite{QOCT}, clocking \cite{clock2,clock}, among many others. It would also allow for wavelength-multiplexing protocols \cite{QIP2} as well as higher dimensional encoding of information \cite{QIP1,QOFC2} to establish quantum networks for information processing and communication. However, the chip-scale photon sources developed so far exhibit fairly limited bandwidths, generally in the order of 100 GHz to a few THz (see for example \cite{Review_photon1,Review_photon2,chip_SPDC,UCSD,chip_SFWM}). To date, broadband entangled photons are only available in bulk devices where a large bandwidth is obtained by a certain spatial modulation of the phase-matching condition, such as chirping the nonlinear grating \cite{chirp1,chirp2,chirp3}, 
cascading nonlinear crystals \cite{multiple}, and spatially modulating the device temperature \cite{refractive}. All these approaches, however, come with a cost of sacrificing the generation efficiency since different frequency components are produced only in different small sections of the device. 
For the same reason, the photons generated may not be transform-limited, requiring post-correction of the spectral phase. 

In this article, we demonstrate generation of bright ultra-broadband entangled photon pairs on a nanophotonic chip using spontaneous parametric down-conversion (SPDC) on the thin-film lithium niobate (LN) platform. This platform has recently attracted significant interest for realizing a variety of nonlinear and electro-optic functionalities \cite{Review_LN1,Review_LN2}. Here we show that specific dispersion engineering of an LN nanowire waveguide is able to offer an extremely broad SPDC phase-matching bandwidth up to 100~THz, with the spectrum covering from 1.2 $\mu$m to 2 $\mu$m, over an order of magnitude larger than a typical chip-scale photon pair source and several times larger the ones with the widest spectrum \cite{Review_photon1,Review_photon2,comp2,comp1,chip_SFWM2}. Together with high-quality periodic poling directly on chip, the device is able to produce photon pair flux as high as $10^{10}$ photons/sec with less than a milliwatt of pump power, with a generation efficiency of 13 GHz/mW that is among the highest ever reported for a broadband entangled photon pair source \cite{disp1,disp2,chirp1,chirp2,chirp3,multiple,comp3,comp4,comp5,comp6,comp1,comp2}. In particular, the produced energy-time entangled photon pairs are of extremely high quality, with coincidence-to-accidental ratio (CAR) in the tens of thousands even for MHz-level pair generation rates. We have obtained an unprecedented CAR of 150,000 within a 17-nm wide spectral region in the telecom band, and a visibility of 98.8\% for Franson-type quantum interference. Even for the entire spectrum, the recorded CAR is as high as 20,000. The high quality of the broadband entangled photons demonstrated here open a path towards many important applications in spectroscopy, metrology, sensing, and information processing that can be realized on a LN photonic integrated circuit.


\begin{figure*}
\centering
  \includegraphics[scale=1]{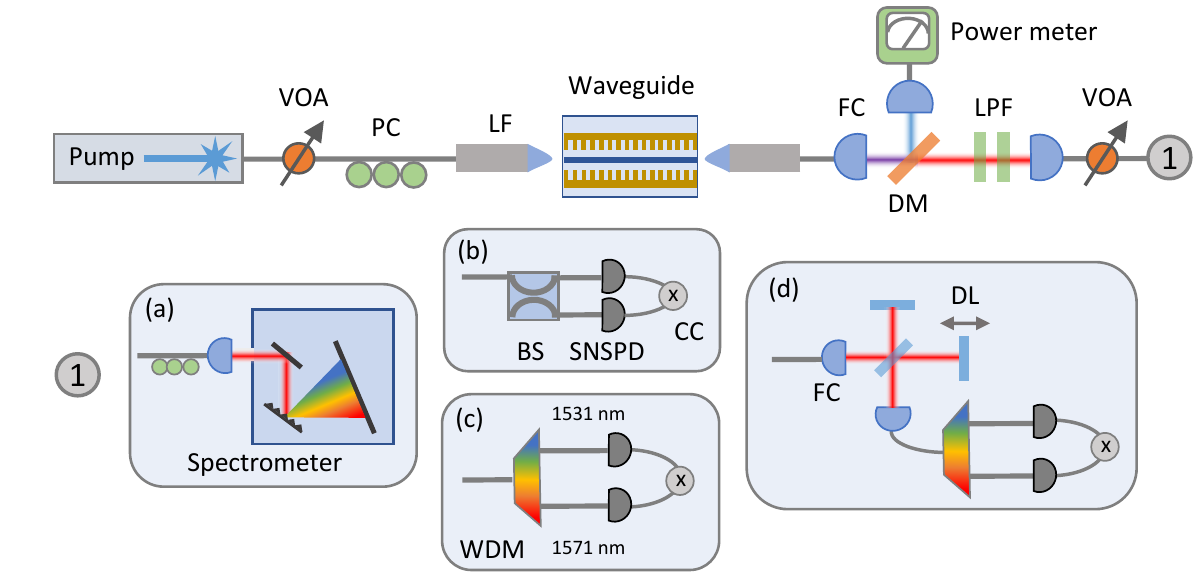}
  \caption{Experimental setup. A pump laser is coupled into the device using a pair of lensed fibers (LF). The generated photon pairs are separated using a dichroic mirror (DM) and a pair of longpass filters (LPF). We run four measurements with the photon pairs: (a) spectrometry, (b) coincidence counting with a beam splitter (BS), (c) coincidence counting of filtered signal and idler photons with a 17 nm bandwidth centred at 1531 nm and 1571 nm using a wavelength-division multiplexer (WDM), (d) Second-order interference of the filtered photons using a Michelson interferometer with an adjustable delay (DL). VOA: variable optical attenuator, PC: polarization controller, FC: fiber collimator, SNSPD: superconducting nanowire single-photon detector, CC: cross-correlator.}
  \label{fig2}
\end{figure*}

\section{Results}
\subsection{Device design and fabrication}
 
Our device generates entangled photons in the fundamental quasi-transverse electric (quasi-TE) mode of a periodically-poled x-cut LN nanowire waveguide (Fig.~\ref{fig1}(a) and (b)), which makes use of the $\mathrm{d_{33}}$ element of LN's $\chi^{(2)}$ tensor with the highest nonlinearity.
The SPDC process relies critically on the phase-matching among the interacting modes to produce a signal-idler spectrum with center frequency $\omega_0$ by a pump photon at frequency 2$\omega_0$. For a waveguide length of $L$, the phase-mismatch $\Delta \varphi = \Delta k L$ is given by 
\begin{align} \label{eq1}
   \Delta \varphi=L[k(2\omega_0)-k(\omega_0-\Delta\omega/2)-k(\omega_0+\Delta\omega/2)- k_{\rm pp}], 
 \end{align}
where $k(\omega)$ is the wavevector of the fundamental quasi-TE mode at a frequency $\omega$ and $\Delta\omega$ is the  frequency difference between the signal and the idler photon. $k_{\rm pp} = \frac{2\pi}{\Lambda}$ is the effective wavevector introduced by the periodic poling with a spatial frequency $\Lambda$. Due to the symmetry of signal and idler frequencies around $\omega_0$ (i.e. $\omega_s+\omega_i = 2\omega_0$) (Fig.~\ref{fig1}(c)), Eq.~(\ref{eq1}) can be written as 
\begin{align} \label{eq1_1}
   \Delta \varphi = L[k(2\omega_0)-2k(\omega_0) - k_{\rm pp}] - 2\sum_{n=1}^\infty{\frac{L \beta_{2n}}{(2n)!}\left(\frac{\Delta\omega}{2}\right)^{2n}}, 
 \end{align}
where $\beta_{2n}=\frac{d^{2n} k}{d\omega^{2n}}$ is the ${\rm 2n}^{\rm th}$-order dispersion of the device. The first term can be eliminated by appropriate periodic poling, $k(2\omega_0)-2k(\omega_0) - k_{\rm pp} =0$. As a result, the phase-mismatch is dominated by the group-velocity dispersion (GVD) $\beta_{2}$ of the device. Therefore, if the GVD of the device is engineered to be zero at $\omega_0$, the phase-matching bandwidth could be significantly broadened, with a phase-mismatch limited only by a small fourth- and higher-order dispersion. On the other hand, the sub-micron cross-section of the waveguide dramatically enhances the nonlinear interaction, resulting in an extremely efficient SPDC process. Consequently, a small waveguide length on the order of a few millimeters is adequate for efficient photon pair generation, which in turn reduces the magnitude of the phase-mismatch considerably (Eq.~(\ref{eq1_1})) and broadens the phase-matching bandwidth even further.


 


The strong confinement of the guided mode in a nanophotonic waveguide enables flexible engineering of the refractive index by introducing geometrically-induced dispersion over the material dispersion. As shown in Fig.~\ref{fig1}(d), by simply changing the width of a LN nanowire waveguide, one is able to flexibly tune the zero-dispersion wavelength in the telecom band. At the same time, the fourth-order dispersion is also engineered to be very small (2.6 $\times$ 10$^{-6}$ ps$^4$/m at 1550 nm) as shown in Fig.~\ref{fig1}(e) which further limits deviations of the phase-mismatch $\Delta\varphi$ from zero. Figure \ref{fig1}(f) shows that the phase-matching bandwidth can reach up to 80 THz for a 5-mm long LN waveguide when the zero dispersion wavelength coincides with the center frequency $\omega_0$.  

 
 
 Using this design, the waveguide is fabricated from 600 nm x-cut lithium niobate thin-film on insulator (LNOI) wafer (see Appendix for details). Figure \ref{figSEM} shows the scanning electron microscope (SEM) images of a fabricated device. The waveguide is subsequently poled using high voltage electrical pulses passing through electrodes patterned on both sides of the waveguide. The poling quality is monitored using the strength of second harmonic generation (see Appendix for details).

\begin{figure}
\centering
  \includegraphics[scale=0.95]{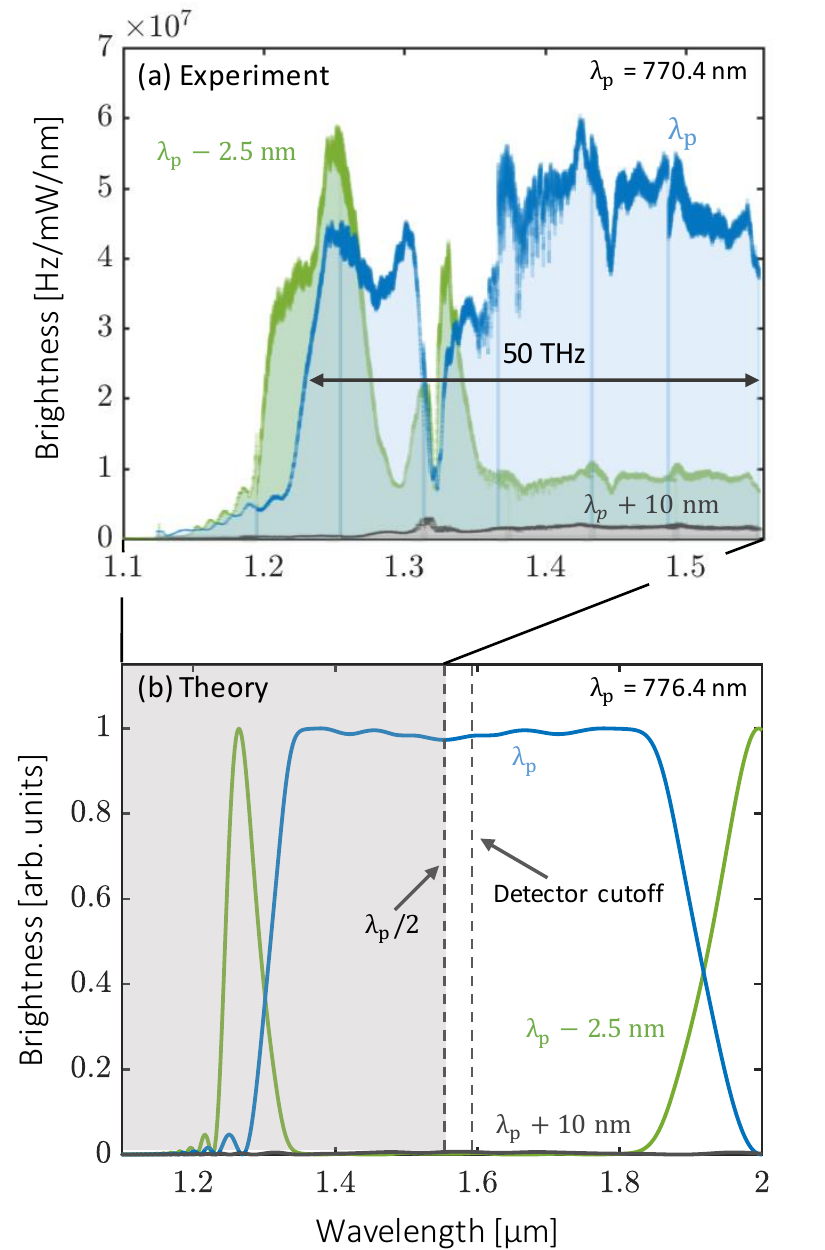}
  \caption{(a) Signal spectrum at different pump wavelengths indicated with each plot with $\mathrm{\lambda_{p}}$ corresponding to the pump wavelength that gives the widest spectrum. (b) The corresponding simulated plots for the designed waveguide.}
  \label{fig5}
\end{figure}

\subsection{Biphoton spectrum} \label{spectrum}
Figure \ref{fig2} shows the experimental setup used to characterize the temporal correlations of the photons as well as their spectrum. A tunable diode laser pumps the waveguide TE mode around 775 nm through a lensed fiber that focuses the light at the waveguide facet. On the other side of the waveguide, the laser and the generated photons are collected and separated using a dichroic mirror and a sequence of two long-pass filters. All together, the pump rejection in the photon pair channel is estimated to be at least 120 dB. We will refer to the photons with wavelength shorter than pump half-wavelength as signal and the photons on the opposite side as idler. The bandwidth of the generated photons is measured directly with an infrared spectrometer with a  dark count rate low enough to measure the generated photons. In this measurement, we scan the pump laser wavelength around 775 nm to evaluate the SPDC bandwidth dependence on pump detuning. The pump wavelength that generates a spectrum with the largest bandwidth in the experiment is 770.4 nm and the corresponding spectrum is plotted in Fig.~\ref{fig5}(a). Here we measure a degenerate SPDC spectrum stretching down to 1200 nm with a 1540.8 nm center wavelength. The 3-dB half-bandwidth evaluates to about 50 THz. The spectrum extends up to the 1590 nm after which the spectrometer's InGaAs CCD cuts off any further measurement at larger wavelengths. Therefore we can not measure the idler side of the spectrum and rely on the signal spectrum to get the bandwidth. Energy conservation dictates that the spectral width of the generated photons should be symmetric around the pump half-wavelength, making the total bandwidth twice that of the measured signal spectrum. This gives a 100 THz (800 nm) total biphoton bandwidth with the idler spectrum expected to span up to 2 $\mu$m in the mid-IR region. To the best of our knowledge, this is the largest SPDC spectral width on any nanophotonic device \cite{Review_photon1,Review_photon2, comp1,comp2,linSPDC}. We compare this result with the state of the art broadband SPDC sources reported to date in Section \ref{disc}. Integrating the spectrum, we get a total on-chip efficiency of 13 GHz/mW of pump power. The high efficiency obtained here is primarily due to the broad phase-matching bandwidth of our device which allows for a large number of frequency modes that can be populated. For an undepleted pump, the photon flux scales with the bandwidth of the photon pairs \cite{flux}. This makes broadband sources ideal for applications in quantum communication and information processing.

\par When the pump is blue-detuned by 2.5 nm, we find a second phase-matching point. Here the photons are strongly non-degenerate in frequency. The signal spectrum has a width of $\sim$ 70 nm  peaked at 1250 nm and the idler is expected to be at 2000 nm. The corresponding theory plots in Fig.~\ref{fig5}(b) show the same phase-matching points with the spectrum switching from degenerate to strongly non-degenerate SPDC for the same blue-detuned pump with good qualitative agreement. The measured spectra at the two phase-matching points and their bandwidths are broader than the simulated ones, most likely due to variations in the waveguide width due to fabrication imprecision and some inhomogeneity in the poling period across the waveguide length which is to be expected. Furthermore, the 6 nm difference between theoretical and experimental phase-matching points as indicated in Fig.~\ref{fig5} is likely due to a small variation in the wafer thickness.

\begin{figure*}
\centering
  \includegraphics[scale=0.98]{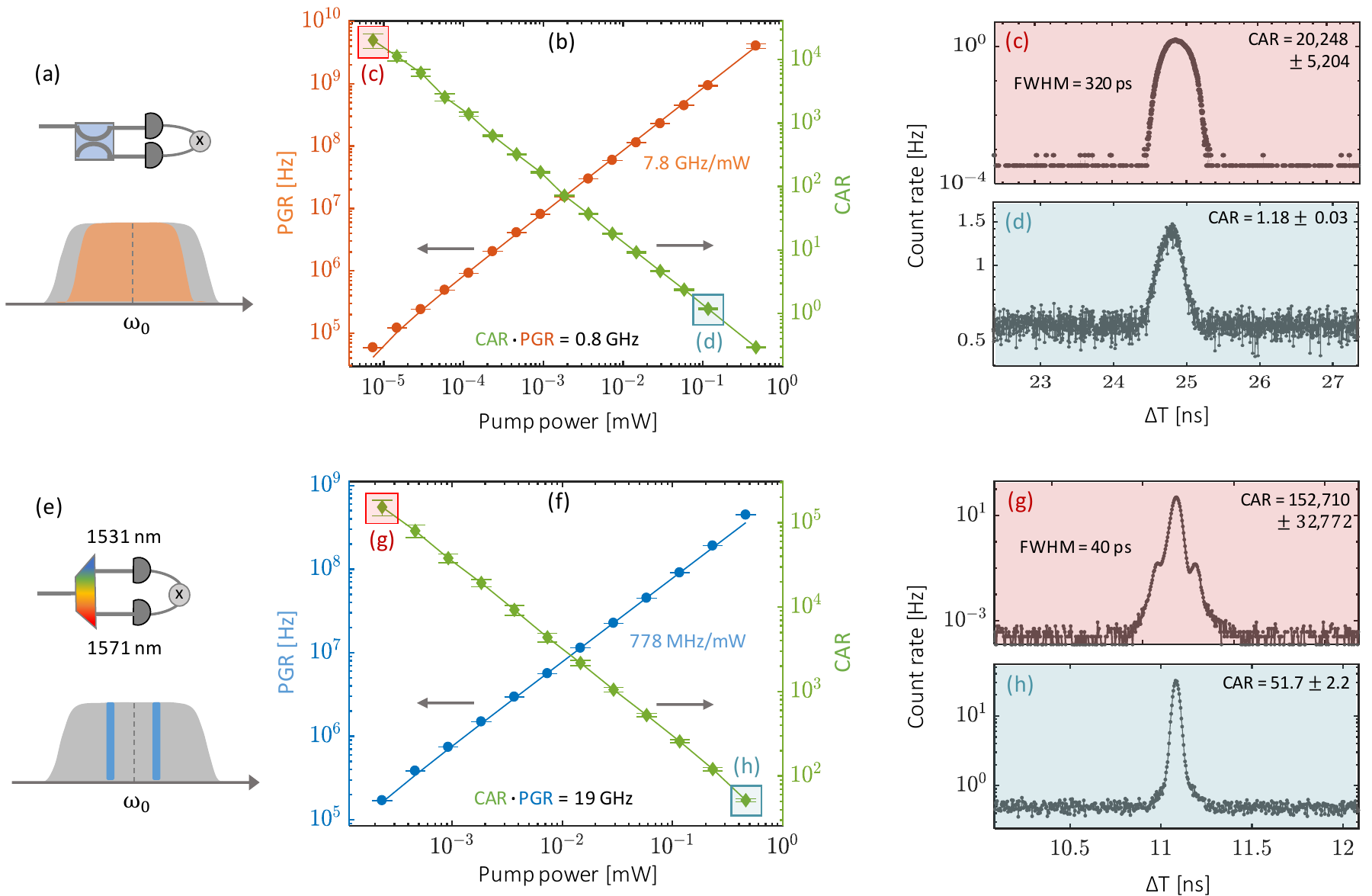}
  \caption{(a) Coincidence measurements for the full spectrum. (b) Pair generation rates and CAR measurement for the photon pairs for increasing pump powers with linear fits. (c) and (d) Coincidence histograms for two points indicated in (b) with the corresponding CAR values. (e) Similar measurements done with the spectrum filtered at 1531 nm and 1571 nm using a wavelength-division multiplexer. (f) The corresponding pair generation rates and CAR measurements and coincidence histograms for two points shown in (g) and (h) indicated in (f). $\mathrm{\Delta}$T: difference in arrival times of signal and idler photons.}
  \label{fig3}
\end{figure*}

\subsection{Temporal correlations}
To characterize the temporal correlations of the generated photon pairs, we split the photons using a 50:50 fiber beam splitter and the two outputs are coupled into superconducting nanowire single-photon detectors (SNSPDs), the outputs of which are fed into a coincidence counter as shown in Fig.~\ref{fig2}(b). This splits the photon pairs across the two outputs of the beam splitter with a 50\% probability. The pair generation rate (PGR), obtained by integrating the coincidence histogram within the biphoton coherence envelope after subtracting the background and accounting for transmission losses, is plotted for increasing pump power in Fig.~\ref{fig3}(b). The data fits cleanly to a straight line as the PGR is linear with the pump power and the generation efficiency can be evaluated as the slope of the line. We obtain an on-chip efficiency of 7.8 GHz/mW. The measured efficiency is limited by the detection bandwidth of the SNSPDs as we obtained roughly twice as high efficiency with the spectrometer. In order to measure the noise characteristics of the source, we evaluate the coincidence-to-accidental ratio (CAR). The results are plotted in green in Fig.~\ref{fig3}(b). The quality of the coincident photons degrades as we increase pump power. This is expected since a higher pump power increases PGR causing more multi-pair generation events. The highest CAR obtained in this experiment is 20,248 $\pm$ 5,204 at a PGR of 52 $\pm$ 0.4 KHz (see Appendix for details on the measurement uncertainties). Although this value is quite high, due to the broad spectrum, the highest achievable CAR is being limited by the experimental setup. For instance the asymmetric wavelength dependence of the SNSPDs' quantum efficiency around the spectral center wavelength, as well as dispersion in the fiber beam splitter causing unequal splitting of the broadband photon pairs, result in increased accidental counts. This is in addition to the 50\% loss in coincidences due to the probabilistic separation of the photons in a pair by the beam splitter. A further decrease in CAR occurs due to dispersion in the optical fibers causing broadening of the coincidence envelope, and increasing the integration window for accidental counts. This can be seen in Fig.~\ref{fig3}(c),(d) where we plot coincidence histograms for the highest and lowest CAR measurements which has a full-width at half-maximum (FWHM) of 320 ps, much larger that the 40 ps timing jitter of the SNSPDs.

\par To demonstrate the true noise characteristics of the generated photons, we filter the signal and idler photons into two 17 nm channels centered at 1531 nm (signal) and 1571 nm (idler) using a telecom band wavelength-division multiplexer (WDM) as shown in Fig.~\ref{fig2}(c). The signal and idler photons are roughly equally spaced from the spectral center when pumped at 775 nm. This removes all wavelength-dependent effects in the measurement process. The results are shown in Fig.~\ref{fig3}(f)-(h). We clearly see in the coincidence histogram in Fig.~\ref{fig3}(g), (h) that the FWHM now shrinks to 40 ps, matching with the timing jitter of the SNSPDs. For this measurement, we obtain a generation efficiency of 778 MHz/mW and a brightness of 38 MHz/mW/nm which is in good agreement with the 43 MHz/mW/nm brightness obtained with the spectrometer in Fig.~\ref{fig5}(a). We obtain our highest CAR of 152,710 $\pm$ 32,772 at a PGR of 176 $\pm$ 0.7 KHz. To the best of our knowledge, this is the highest CAR obtained on a chip-scale device \cite{comp2,CAR1,CAR2,CAR3,CAR5,CAR6,CAR7,CAR8,CAR9,CAR10,UCSD}. An even higher CAR can actually be obtained using smaller pump powers and integrating for longer periods of time \cite{maxCAR}, ultimately limited by dark counts of the photodetectors for SPDC sources. In our experiment, the highest CAR was obtained at a signal-to-noise ratio of 20 between the singles counts and dark counts at the detectors indicating that we can still obtain a higher CAR by reducing the pump power further. A better estimate of the noise characteristics of a photon pair source is the product of CAR and PGR obtained at higher optical powers which evaluates device performance independently of the dark counts and pump power in the experiment. The product of CAR and PGR evaluates to 19 GHz and 0.8 GHz for the filtered and full spectrum respectively, indicating low noise even at high rates of photon emission. The high performance of the generated photons on WDM-filtered channels demonstrates their capability for wavelength-multiplexed quantum communication in telecom-band fibers \cite{QIP2}.

\begin{figure}
\centering
  \includegraphics[scale=1]{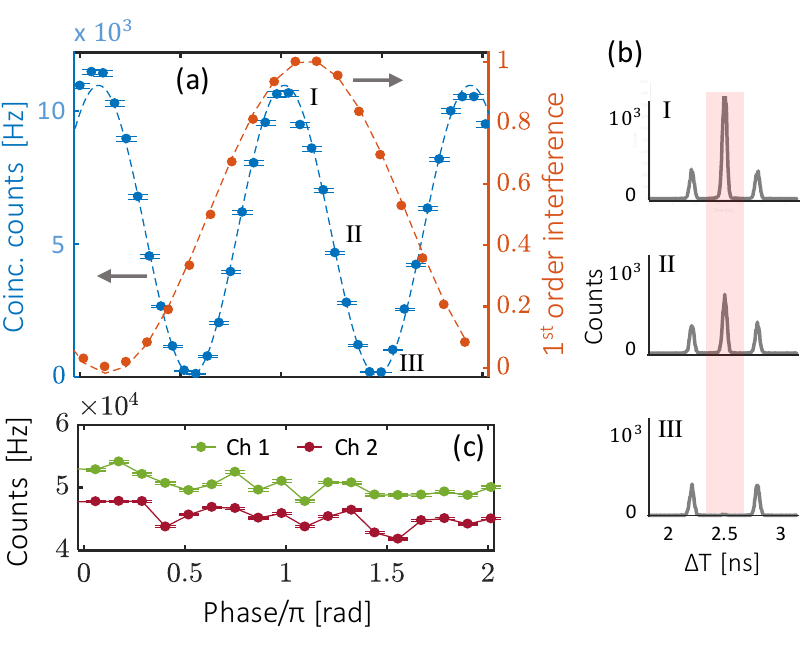}
  \caption{(a) Two-photon interferogram for the filtered spectrum with sinusoidal fit, along with first-order interference with a laser at a similar wavelength as the photons. (b) Sample coincidence histograms versus signal-idler delay $\mathrm{\Delta T}$ at three points indicated in (a). The central peak in the coincidence histogram oscillates with the interferometer phase and is filtered and integrated to obtain the interferogram. (c) The corresponding singles counts at the two detectors indicating no first-order interference in the photon pairs.}
  \label{fig4}
\end{figure}

\par In order to verify the non-classical behavior of the generated light, we run a two-photon interference experiment. The time-energy entanglement present in SPDC can be verified using a Bell's inequality violation \cite{bell} by beating a 70.7\% visibility limit in two-photon interference for. For this, we run an interference experiment in a folded Michelson interferometer \cite{Michelson} on the filtered signal and idler photons as shown in Fig.~\ref{fig2}(d). This will verify the time-energy entanglement in the photon pairs and demonstrate the enhanced phase sensitivity of non-classical light sources, without making any conclusions on the nonlocal nature of entanglement due to the shared optical path between the two photons. When the signal and idler photons are inserted into the interferometer, there are four possible paths for the two photons to take. If the signal takes the longer path and the idler takes the shorter path, or vice versa, the two outcomes become distinguishable due to the additional delay incurred by one of the photons which reveals the path information, prohibiting any interference. This presents as two distinct peaks in the coincidence histogram delayed by twice the optical path length in the interferometer as shown in Fig.~\ref{fig4}(b). However, if both photons take either the shorter or the longer path, the two outcomes are indistinguishable leading to interference between them. These two outcomes create a third peak with an amplitude that oscillates with the phase difference in the interferometer as shown in Fig.~\ref{fig4}(b). We integrate coincidence counts within the envelope of this peak to obtain the interferogram. Figure \ref{fig4}(a) shows the results of the measurement. For each phase we integrated for 30 seconds to obtain a reasonable number of counts at a PGR of 380 KHz. The interference data is fitted to a sinusoid giving a visibility of 98.8\%, violating the Bell's inequality. We further compare this result with a first-order interference done with a laser as shown in orange in Fig.~\ref{fig4}(a) demonstrating halving of the interference period for the photon pairs. This measurement was not done for the full spectrum due to its large bandwidth which will cause significant experimental challenges such as dispersion in the interferometer and the optical fibers which will artificially reduce the visibility, and the absence of filters that can be tuned over the 800 nm bandwidth. We expect similar performance for the rest of the spectrum.

\section{Discussion} \label{disc}
\begin{figure}
\centering
  \includegraphics[scale=0.95]{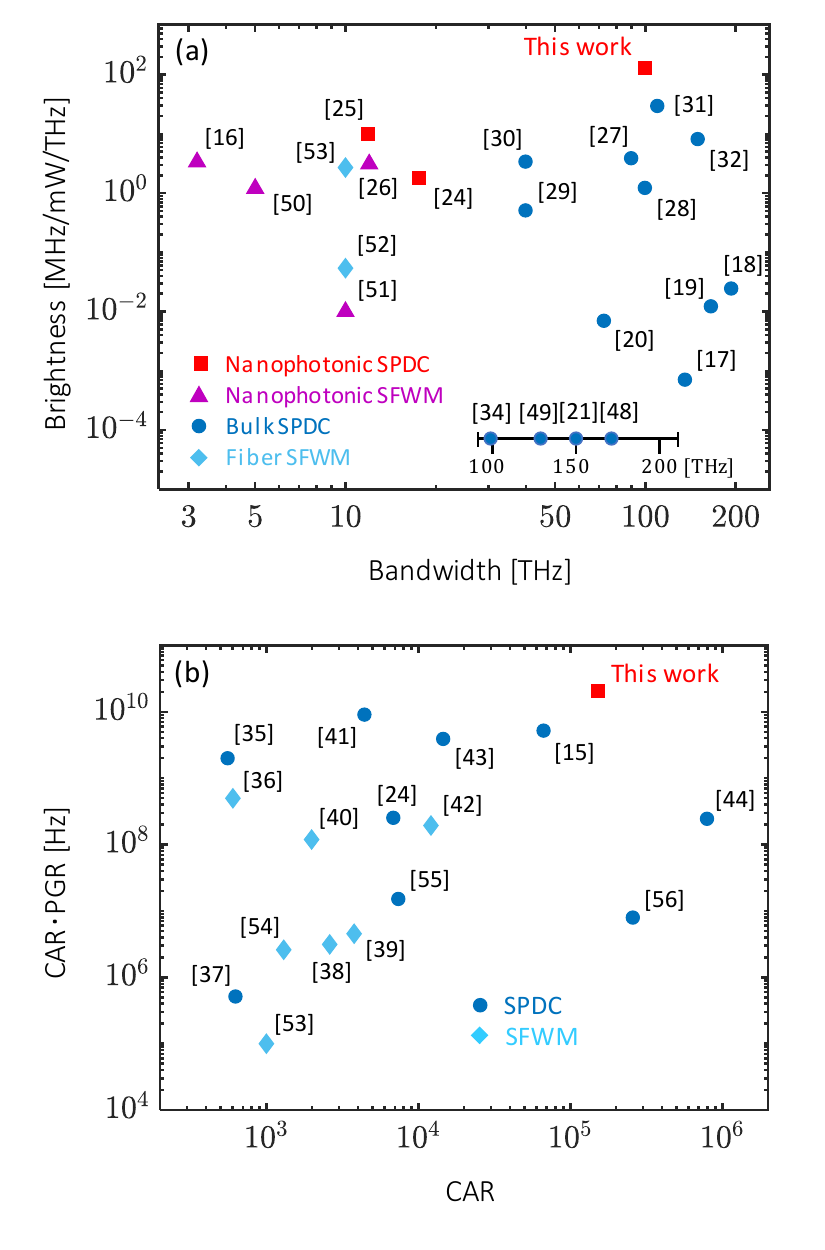}
  \caption{(a) Bandwidth and brightness for some recent demonstrations of broadband SPDC and SFWM. Inset at the bottom shows references for which brightness is not available. Brightness was evaluated from reported photon flux, input power and number of modes in \cite{comp4}, from reported photon flux, and laser power in \cite{comp3,disp2}, from ratio of singles flux to coincidence flux and reported laser power in \cite{chirp3,multiple}, from plotted photon flux, filter width and input power in \cite{chirp2,CAR4}, from laser pulse width and peak power, photons per pulse and DWDM channel width in \cite{chip_SFWM2}. Bandwidth was extracted from plotted spectrum in \cite{comp3,disp1,disp2,chirp1,comp7} and WDM bandwidth in \cite{chip_SFWM}. (b) Maximum CAR and CAR PGR product for for some recent demonstrations of SFWM and SPDC. PGR was evaluated from stated coincidence rate in \cite{CAR4}, from photon pairs per pulse and laser rep. rate in \cite{CAR5,CAR8,maxCAR}, from stated brightness, bandwidth and input power in \cite{CAR7,comp2}, extracted from PGR plot in \cite{max-1CAR}.}
  \label{fig6}
\end{figure}

Efforts to generate broadband entangled photons over the last two decades have primarily been motivated by applications in metrology. Specifically in biological imaging, broadband entangled photon pairs combine the short correlation time associated with ultra-short laser pulses and the spectral resolution of narrowband light due to the monochromatic pump used to drive the parametric interaction \cite{pulse}. Furthermore, entangled photons increase two-photon absorption probability without a need to increase the intensity of light to dangerous levels \cite{TPAmicroscope}, ideal for nonlinear microscopy. To that end, research into generating photon pairs with coherence lengths approaching an optical cycle have continued. In Fig.~\ref{fig6}(a), we compare the bandwidth and efficiency of some of the state-of-the art broadband SPDC sources to our device. Most of these sources are bulk crystals and waveguides carved into bulk wafers since there has not been much work done on nanophotonic SPDC sources for bandwidth. The plot demonstrates the superior performance of our device, requiring much lower optical powers than others due to the wavelength-scale confinement of light in the nanophotonic waveguide which enhances the light intensity and the nonlinear coupling compared to bulk optics. We note that this list is in no ways complete but a good representative of the work done so far. There have been numerous other demonstrations of similar bandwidths, but a lot of these experiments do not report the device efficiency since that is not their primary concern. We reference some of these results \cite{flux,refractive,comp7,comp8} in the inset of Fig.~\ref{fig6}(a). We also include results obtained with spontaneous four-wave mixing (SFWM) in semiconductor waveguides and optical fibers \cite{chip_SFWM,chip_SFWM2,chip_SFWM3,chip_SFWM4,fiber_SFWM1,CAR4}. Since SFWM is a third order nonlinear process with a photon flux quadratic with pump power, we compare the brightness of these sources at a pump power of 1 mW. The comparison demonstrates that waveguide dispersion can greatly enhance bandwidth of photon pairs in SPDC rather effortlessly. This can be seen by the fact that a different geometry of this waveguide gave a bandwidth of $\sim$ 1 nm for the photon pairs with a similar center wavelength in a previous experiment \cite{UCSD}. This simplifies the design of broadband SPDC sources to a parameter optimization problem of waveguide dimensions to get a desired bandwidth instead of a complicated fabrication process to control modulation of poling period or the refractive index. It is possible to further optimize the device design in our experiment to increase the phase-matching bandwidth by correcting for higher order dispersion and relaxing pump wavelength constraints, a topic for future investigations.

\par In Figure \ref{fig6}(b) we compare the noise performance of our device with the state-of-the-art SPDC and SFWM sources. Here we plot the product of maximum CAR and PGR for a few recent experiments that have reported among the highest CAR values ($>$ 500) \cite{CAR1,CAR2,CAR3,CAR4,CAR5,CAR6,CAR7,CAR8,CAR9,CAR10,CAR11,CAR12,maxCAR,max-1CAR,UCSD}. As discussed earlier, a higher CAR can be obtained with better detectors and lower pump powers for the same device. This indicates that the highest reported CAR can not always be associated with superior device performance, especially for SPDC sources since there is negligible Raman or Brillouin noise background. A CAR $\cdot$ PGR product gets rid of the dependence on experimental setup giving a good estimate of device performance. Our device gives among the highest reported CAR $\cdot$ PGR product, noting that much higher CAR values have been reported albeit at much lower coincidence rates \cite{maxCAR,max-1CAR}. 

\section{Conclusion}
To summarize,  we have presented a broadband photon pair source based on thin-film lithium niobate. Our device uses periodic poling with a fixed period and employs waveguide dispersion to create a broad phase-matching bandwidth. The generated photons have a spectral width of 100 THz, the highest reported for any nanophotonic photon pair source. The spectrum covers wavelengths from 1.2 $\mu$m to over 2 $\mu$m. Due to the broad spectral width, the generated photon flux as high as 13 GHz with a milliwatt of pump power can be obtained, making it highly efficient. Time domain measurements reveal high degree of correlation and excellent noise performance with the coincidence-to-accidental ratio exceeding ten thousand even for MHz-level pair generation rates. The generated photons are highly spectrally correlated as shown by a near-unity quantum interference visibility. Additionally, the electro-optic and piezo-electric effect in LN can allow control of the bi-photon spectral width and its correlations \cite{javid} as well as design of reconfigurable photonic circuits for on-chip manipulating and multiplexing. These qualities make a strong case for nanophotonic LN devices as good sources for wavelength-multiplexed quantum communication and and entanglement distribution. Furthermore, we envision that this work will motivate efforts to bring femtosecond metrology, spectroscopy and nonlinear microscopy to nanophotonic platforms.

\appendix
\section{Device Fabrication and poling}
The device was fabricated on a 600 nm-thick x-cut single-crystalline LN thin film bonded on a 4.7-$\mu$m silicon dioxide layer and a silicon substrate (from NanoLN). The waveguide was first patterned with ZEP-520A positive resist via e-beam lithography, which was then transferred to the LN layer by 300 nm etching via Ar$^+$ ion milling process. The resist residue was removed by an  O$_2$ plasma cleaning process. Next, a second exposure is performed to define the comb-like electrode structures on ZEP-520A resist using e-beam lithography. The electrodes (10 nm Ti/400 nm Au) were finally deposited via e-gun evaporating followed by a standard lift-off process. After fabrication, the waveguide is poled using a sequence of 280 V 10 ms square-wave electrical pulses. The length of the poled region of the waveguide is 5 mm. The poling efficiency is monitored heuristically by coupling a tunable telecom-band laser and measuring the strength of the second harmonic frequency after each pulse. For this device, the second harmonic strength was saturated after applying about 35 pulses.

\section{Data Acquisition}
The temporal correlation measurements are done using free-running SNSPDs connected to a coincidence counter. For measurements of pair generation rates, the coincidence histogram was integrated for 300 seconds for each data point for the full spectrum and 140 seconds for the filtered spectrum. to obtain PGR, the counts were integrated within a width of roughly three standard deviations of Gaussian coincidence envelope, as integrating within the FWHM only would ignore a lot of true coincidences. The obtained PGR values were scaled by the transmission losses of the experimental setup. This consisted of a chip to fiber out-coupling loss of 6.9 dB and 3.2 dB of loss for the rest of the experimental setup from the lensed fibers to the input of the SNSPDs which included the filters, dichroic mirror, fiber collimators and a beam splitter. An additional 1.3 dB loss was incurred by the WDM for the filtered spectrum. All loss corrections took into account the quantum efficiency of the two SNSPDs which were 53\% and 54\% as well as the 50\% probability of separating the two photons at the two outputs of the beam splitter for the pair generation rate calculation. An additional variable fiber attenuator (insertion loss: 2.5 dB) was used before the SNSPDs at higher pump powers since the photon fluxes were high enough to saturate the detectors. 
\par The coincidence-to-accidental ratio (CAR) is defined as CAR = $\mathrm{(C-A)/A}$, where $\mathrm{C}$ represents the the coincidence counts obtained by integrating the counts in a fixed time window within the bi-photon coherence envelope and the accidental counts $\mathrm{A}$ are obtained by integrating within the same window away from the coherence envelope to obtain the noise counts. We set the integration window to be the FWHM. For CAR measurements, the acquisition time was steadily increased from 140 seconds for filtered spectrum and 300 seconds for the full spectrum at high pump powers up to 2 hours at the lowest pump power for the highest CAR value. Increasing acquisition time at lower optical powers was necessary to reduce the uncertainty of the noise floor. For the CAR $\cdot$ PGR product, the PGR was scaled by 76\% as this is the proportion of the area within the FWHM of a Gaussian to its total area.
\par The spectrometer measurements were done from 1150 nm to 1550 nm in 70 nm pieces as this is the maximum bandwidth allowed by the spectrometer's CCD grid. The diffraction grating was rotated by a fixed angle after each acquisition to get to the adjacent wavelengths. Each acquisition was 2 min long. The final spectral plots in Fig.~\ref{fig5} are obtained by stitching the pieces together after subtracting background and accounting for all transmission losses. Small discontinuities between the pieces are primarily due to the slight misalignment in the nano-positioning stages which control the lensed fibers that couple the light in and out of the waveguide.  

\section{Measurement Uncertainties}
The uncertainty in a coincidence counting measurement with counts $\mathrm{C}$ obtained by integrating the coincidence histogram within a fixed time window is given by $\mathrm{E_c=\sqrt{C}}$ using Poisson statistics. The noise floor gives additional accidental counts $\mathrm{A}$ within the same time window. The standard deviation in $\mathrm{A}$ given by $\mathrm{E_A}$ is obtained by creating an ensemble of accidental counts by integrating the noise floor counts within the same time window at different time delays on the histogram. We use the full 262 ns time delay range allowed by the coincidence counter as this gives an ensemble of thousands of accidental count instances $\mathrm{A_j}$. The mean and standard deviation of this ensemble gives $\mathrm{A}$ and $\mathrm{E_A}$, respectively. The PGR is then given by PGR = $\mathrm{C-A}$ with an error given by $\mathrm{\sqrt{E^2_C+E^2_A}}$. The error bars for the two-photon interference experiment are obtained in a similar way. The uncertainty in CAR = $\mathrm{(C-A)/A}$ is given by $\mathrm{\frac{E_{CAR}}{CAR}=\sqrt{(\frac{E_c}{C})^2+ (\frac{E_A}{A})^2}}$. The errors in singles counts are obtained by integrating the singles counts within 100 ms time bins over a period of 10 - 15 seconds on one of the SNSPDs and calculating the standard deviation of the count rate.

\begin{acknowledgments}
This work is supported in part by National Science Foundation (NSF) (EFMA-1641099, ECCS-1810169, and ECCS-1842691); the Defense Threat Reduction Agency-Joint Science and Technology Office for Chemical and Biological Defense (grant No. HDTRA11810047); and the Defense Advanced Research Projects Agency (DARPA) under Agreement No. HR00112090012. This work was performed in part at the Cornell NanoScale Facility, a member of the National Nanotechnology Coordinated Infrastructure (NNCI), which is supported by the National Science Foundation (Grant NNCI-2025233).
\end{acknowledgments}


\end{document}